\documentclass{cernyrep}
\usepackage{texnames}
\usepackage[T1]{fontenc}
\def\cnub{C$\nu$B}
\newcommand{\lwig}{\mbox{\;\raisebox{.3ex}
    {$<$}$\!\!\!\!\!$\raisebox{-.9ex}{$\sim$}\;}}

\begin{document}
\title{The connection between cosmology and neutrino physics}

\author{Steen Hannestad}

\institute{Department of Physics and Astronomy, University of Aarhus, DK-8000 Aarhus C, Denmark}

\maketitle 

\begin{abstract}
Cosmology provides a unique and very powerful laboratory for testing neutrino physics. Here, I review the current status of cosmological neutrino measurements. Future prospects are also discussed, with particular emphasis on the interplay with experimental neutrino physics. Finally I discuss the possibility of a direct detection of the cosmic neutrino background and its associated anisotropy.
\end{abstract}

\section{Introduction}

Neutrino physics provides one of the prime examples of the interplay between particle physics and cosmology.
Because neutrinos are so abundant, the cosmic neutrino background contributes significantly to the cosmic energy density at all times, and therefore have a profound influence on the evolution of our Universe. At early times, around the epoch of neutrino decoupling at $T \sim 1$ MeV,
they influence the formation of light nuclei, and later they have an influence on cosmic structure formation.
This also means that precision cosmology can be used to probe details of neutrino physics, such as the absolute
value of neutrino masses, and the presence of light or heavy sterile neutrinos.

Here I will focus on the use of cosmology to probe the mass of neutrinos. As it turns out the main influence of light neutrinos on structure formation comes via their contribution to the energy density. At late times when at least some of the mass eigenstates are non-relativistic, the energy density in neutrinos can be quantified via the sum of neutrino masses, $\sum m_\nu$. To a very good approximation the effect of neutrinos on structure formation can be described using just this one parameter, i.e.\ it does not matter how the mass is distributed internally between the different states (see \cite{Lesgourgues:2004ps,Lesgourgues:2006nd}
for a thorough discussion of this).

Interestingly, cosmology is sensitive to a different combination of mass eigenstates than other experimental probes.
In neutrinoless double beta
decay experiments, the important parameter is the coherent sum \cite{Aalseth:2004hb,Bilenky:1987ty}
\begin{equation}
m_{\beta\beta} = \left|c_{13}^2 c_{12}^2 m_1 + c_{13}^2 s_{12}^2 m_2
e^{i \phi_2} + s_{13}^2 m_3 e^{i \phi_3} \right|,
\end{equation}
which allows for phase cancellation. The current best upper bound on $m_{\beta \beta}$ comes from the Heidelberg-Moscow experiment and is $m_{\beta \beta} < 0.27 \, {\rm eV}$ (90\% C.L.) \cite{KlapdorKleingrothaus:2000sn,Rodin:2007fz}.

However, the most direct current upper bound on the neutrino mass comes from the final data analysis of the Mainz experiment, and yields $m_{\nu_\beta} \leq 2.3$ eV at 95\% C.L. \cite{kraus}, where
\begin{equation}
m_{\nu_\beta} = \left(c_{13}^2 c_{12}^2 m_1^2 + c_{13}^2 s_{12}^2 m_2^2 +
s_{13}^2 m_3^2 \right)^{1/2}
\end{equation}
is effective parameter measured in beta decay spectra. In this case the mass states are weighed with their mixing with $\nu_e$.

Assuming just the three active neutrino species this corresponds to an upper bound on the sum of neutrino masses of $\sim 7$ eV. The current neutrino temperature is $T_{\nu,0} \sim 1.7 \times 10^{-4}$ eV so that any mass eigenstate heavier than this is non-relativistic at present. The contribution of neutrinos of any mass eigenstate, $i$, to the current energy density is given by $\Omega_{\nu,i} h^2 = m_{\nu,i}/93$ eV, where $\Omega$ is the density parameter and $h$ the Hubble parameter in units of $100 \, {\rm km} \, {\rm s}^{-1}$. The total neutrino contribution to the energy density is therefore $\Omega_{\nu,i} h^2 = \sum_i m_{\nu,i}/93$ eV, where the sum is over all non-relativistic states.

The current upper bound on the dark matter density is roughly $\Omega_{\rm dm} \lwig 0.1$ so that for $\sum m_\nu$, neutrinos would make up a very large fraction of the dark matter density.
However, this is strongly ruled out by observations because of the free streaming property of neutrinos. Being light particles, neutrinos are relativistic approximately until the epoch of matter radiation equality. This means that all neutrino structures inside the horizon at this epoch have been erased, and if neutrinos constituted all the dark matter structure formation would have been impossible.
This possibility is clearly excluded, and the argument can be refined to set constraints on the neutrino mass using precision measurements of cosmic structure formation.

\section{Current constraints on neutrino properties}

The most important data set currently used to constrain cosmology is the measurement of the CMB anisotropy by the WMAP experiment \cite{Komatsu:2008hk}.
However, some parameters do not have a significant impact on the CMB and must be constrained by adding other data. One example is the
neutrino mass which leads to a suppression of fluctuation power on scales smaller than the free-streaming scale. In linear theory the
relative change in the matter power spectrum is roughly
\begin{equation}
\frac{\Delta P}{P} \sim -8 f_\nu,
\end{equation}
with $f_\nu = \Omega_\nu/\Omega_m$. The transition occurs smoothly around the free-streaming scale, given
very approximately by \cite{MB}
\begin{equation}
k_{\rm FS} \sim 0.8 \frac{m_\nu}{\rm eV} \, h \, {\rm Mpc}^{-1}
\end{equation}
A detailed discussion of these issues can be found in \cite{Lesgourgues:2006nd}.
This damping of power is best constrained by either large scale structure (LSS) surveys which directly measure the spectral shape of the matter power
spectrum, or by measurements of the fluctuation power amplitude on small scales. The LSS surveys like SDSS are generally more robust because they do not explicitly rely on measurements in the very non-linear regime, but formally stronger constraints can be obtained by using data from the Lyman-$\alpha$ forest or from the cluster mass function.

A very large number of papers have been dedicated to the study of how neutrino properties can be constrained using such data
(see \cite{Komatsu:2008hk,numass,reid} for an incomplete list). Most studies have focussed on two effective parameters, $\sum m_\nu$, and the effective number of neutrino species, $N_\nu$. The latter parameter can be defined in a number of way, but the most widely used definition in this context is to have the total neutrino mass, $\sum m_\nu$ distributed among $N_\nu$ species of equal mass. The standard model predicts $N_\nu = 3.04$ with the 0.04 coming from incomplete neutrino freeze-out and finite temperature effects around $e^+ e^-$ annihilation \cite{decoupling}. However, this number can be changed by the presence of sterile neutrinos, a finite neutrino chemical potential or other light particles such as axions.

Here I show an example calculation using the latest cosmological data: CMB measurements from WMAP-5 \cite{Komatsu:2008hk}, as well as the matter power spectrum measured from the SDSS-DR7 LRG data \cite{reid}. In addition I add a prior from the latest estimate of the Hubble parameter \cite{riess}.

Another important issue in cosmological parameter estimation is the complexity of the model parameter space. Most studies have focussed on neutrino constraints on top of the simplest vanilla $\Lambda$CDM model such that {\it either} $\sum m_\nu$ or $N_\nu$ is fitted in addition to the usual cosmological parameters: $\Omega_m$, the matter density, $\Omega_b$, the baryon density, $H_0$, the Hubble parameter, $n_s$, the scalar spectral index, $A_s$, the scalar amplitude, and $\tau$, the optical depth to reionization. However, this approach may be too naive, it could well be that other physics like dark energy with a non-trivial equation of state, $w$, has to be added together with neutrino physics in the fit. Some papers have discussed the possible degeneracy between neutrino parameters and other cosmological parameters.

\begin{figure}[h!]
\hspace{25mm}
\includegraphics[width=0.8\linewidth]{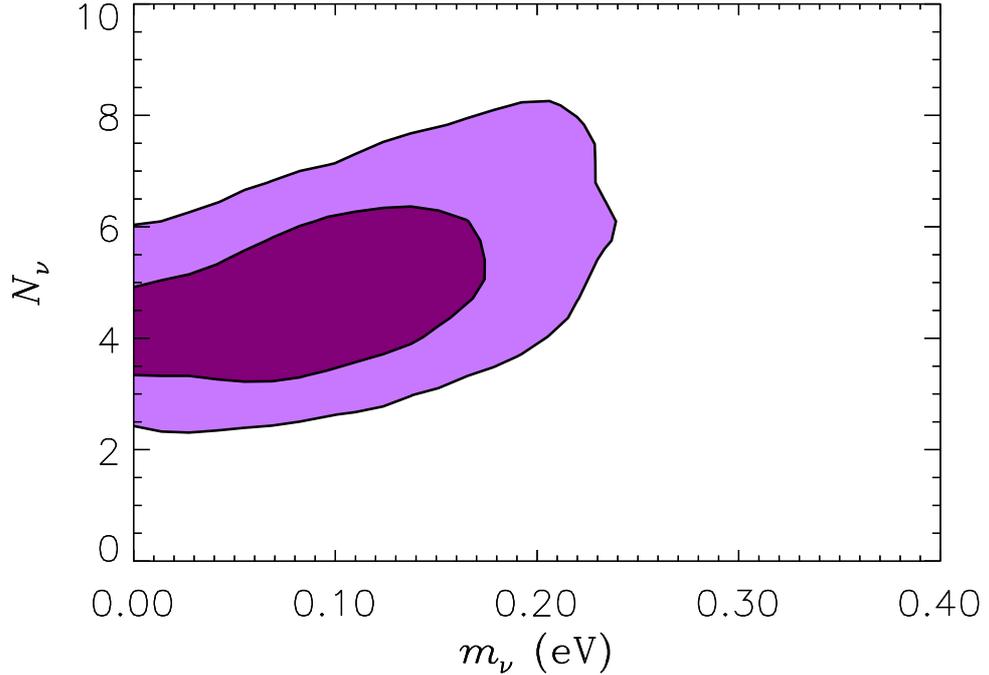}
\caption{68\% and 95\% contours in the $m_\nu$-$N_\nu$
  plane using WMAP-5, SDSS-DR7 and $H_0$ data.
\label{fig:contours}}
\end{figure}

In Fig.~\ref{fig:contours} I show the contraint on $m_\nu = \sum m_\nu/N_\nu$ and $N_\nu$ from WMAP-5, SDSS-DR7 LRG, and $H_0$ data. The mass constraint on each mass state is $m_\nu \lwig 0.2$ eV at $2\sigma$, compatible with other recent analyses. It should again be stressed that the bound can be strengthened by adding other data sets which are, however, less robust. The preferred value of $N_\nu$ is larger than the value predicted by the standard model, again in accordance with other recent findings. However, $N_\nu=3.04$ is within the 95\% contour which is $3.03 < N_\nu < 7.59$ and even though the result is interesting and could be pointing to the presence of for example a new light sterile state of small mass, the statistical significance is very weak.

\begin{figure}[h!]
   \noindent
   \begin{minipage}{0.33\linewidth}
     \hspace*{-0.5cm}\includegraphics[width=1.02\linewidth]{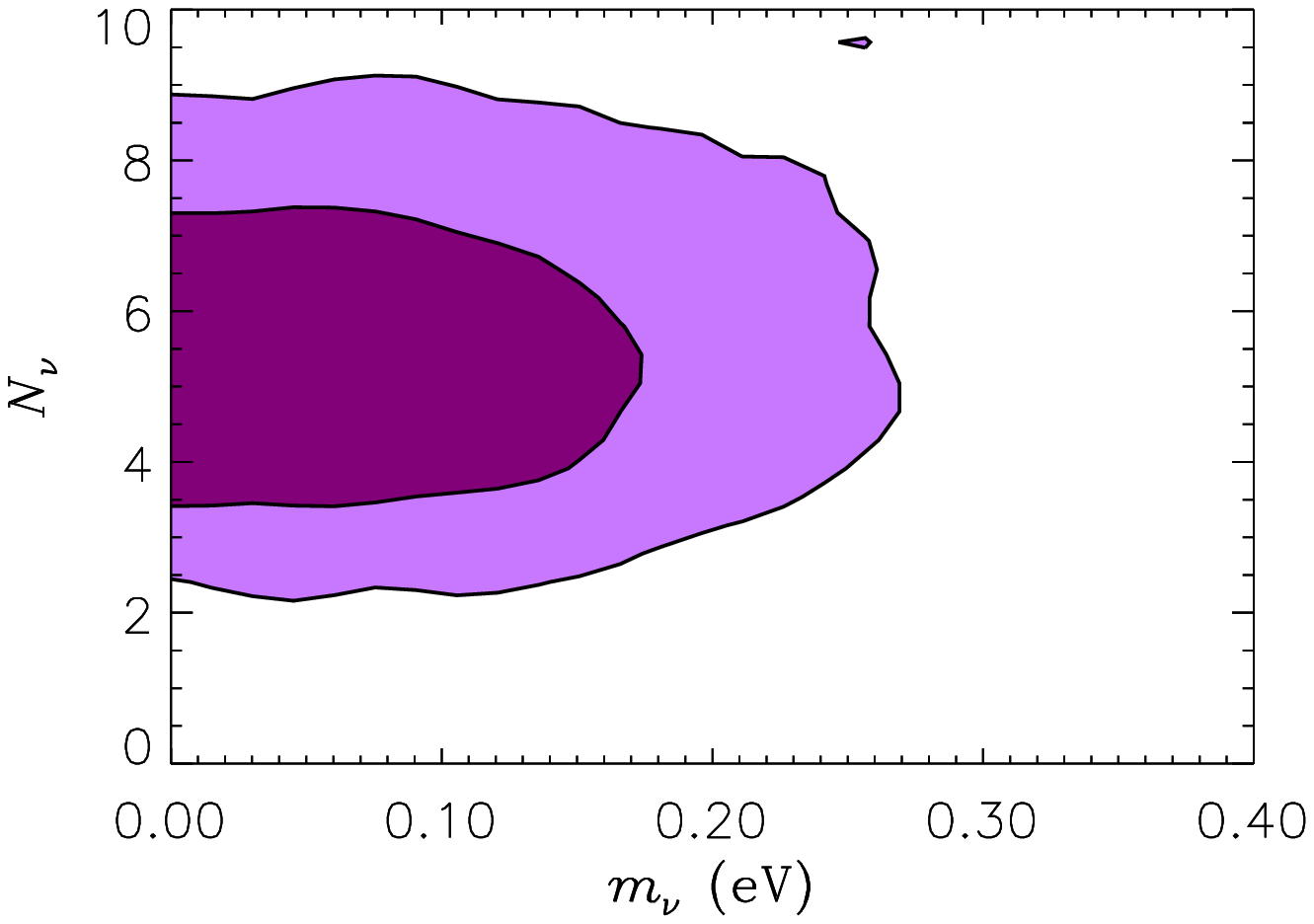}
   \end{minipage}
   \begin{minipage}{0.33\linewidth}
     \hspace*{-0.2cm}\includegraphics[width=1.0\linewidth]{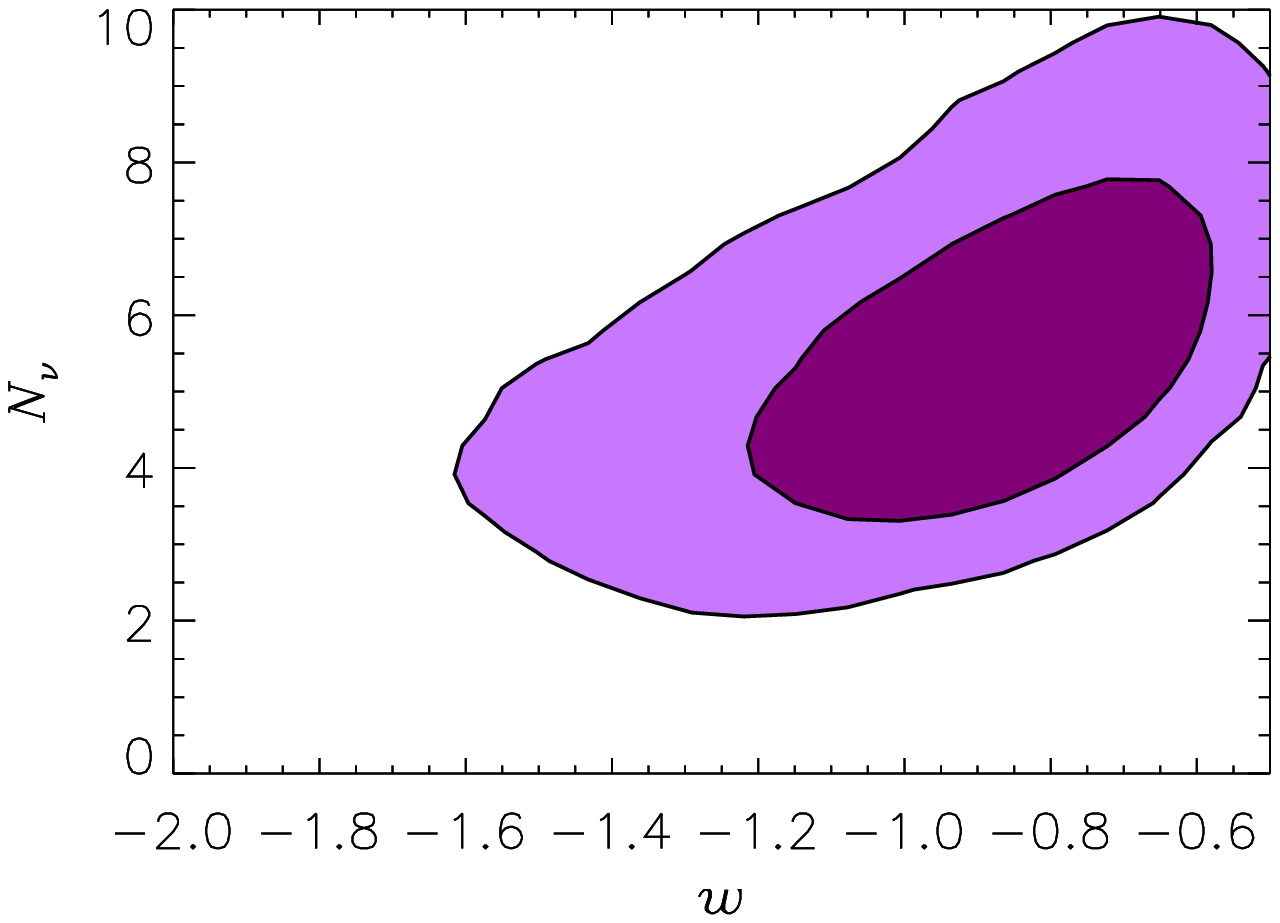}
   \end{minipage}
   \begin{minipage}{0.33\linewidth}
     \hspace*{0.2cm}\includegraphics[width=1.06\linewidth]{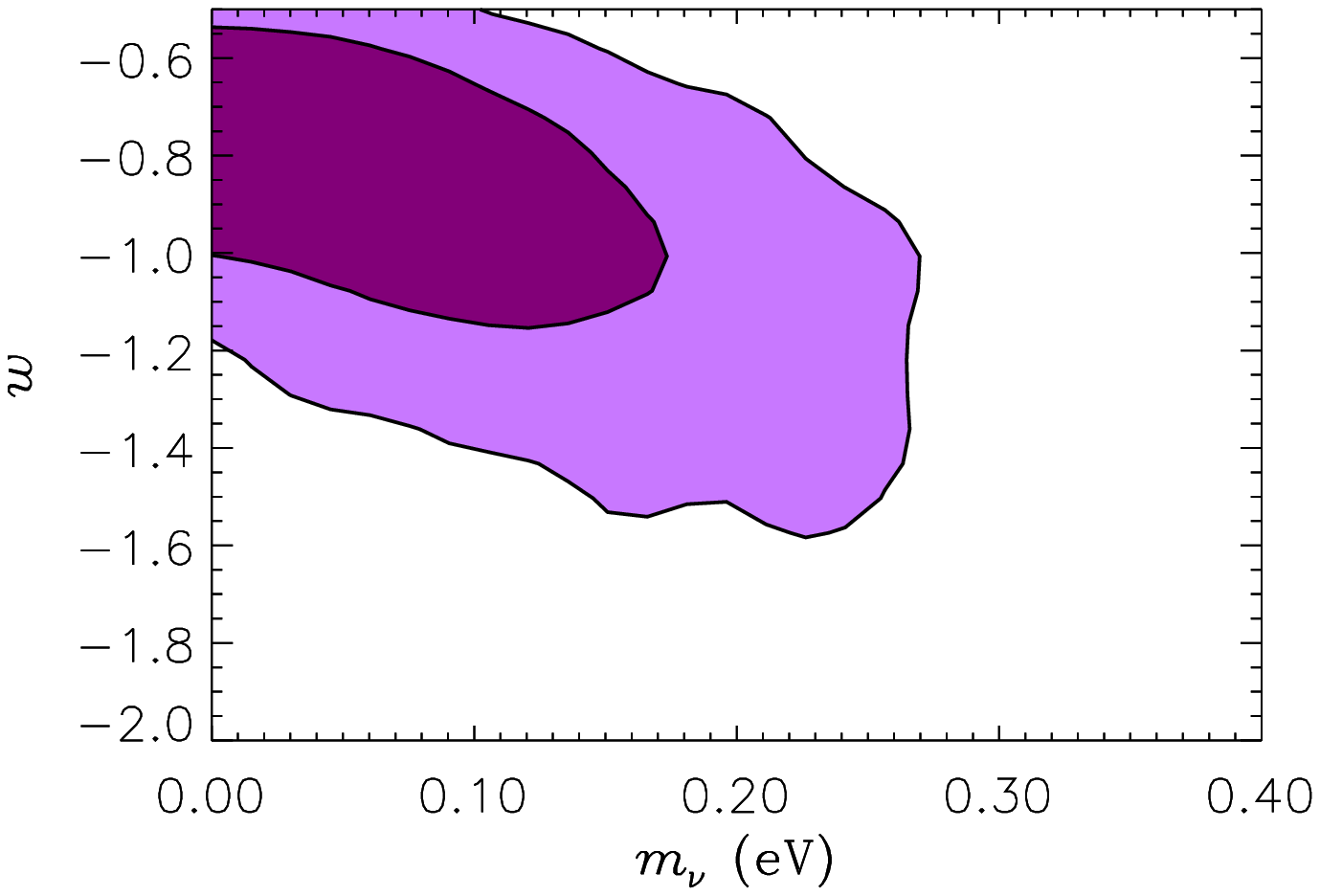}
   \end{minipage}
\caption{68\% and 95\% contours for $m_\nu$, $N_\nu$, and $w$
   using WMAP5, SDSS-DR7 and $H_0$ data.
\label{fig:w}}
\end{figure}

In Fig.~\ref{fig:w} I show an analysis with the same data, but now adding the equation of state of dark energy, $w$, as a free parameter.
The figure shows a significant degeneracy between $m_\nu$, $N_\nu$, and $w$. Even though there is a degeneracy the formal bound on $m_\nu$ does not change much. The same is true for the lower bound on $N_\nu$ which is 2.97 at 95\% C.L., but the upper bound is relaxed significantly from 7.59 to 8.70.

\section{Future constraints on neutrino properties}

New and much more precise CMB measurements will be provided by the Planck satellite, launched in May 2009 {\bf CHANGE}.
In the coming years there will also be a large number of new cosmological experiments designed for precision measurements of structure formation.
There are two ways of improving current data, either measuring larger effective volumes, or going to higher redshifts where structures are more linear. In practise many of the proposed experiments will do both. One example is measurements of weak gravitational lensing by the Large Synoptic Survey Telescope (LSST) which will have an effective volume many times larger than the SDSS and measure out to redshifts of order 1.5 \cite{lsst}.

One very powerful way to discriminate massive neutrinos from other physical effects acting to suppress fluctuations on small scales is to look at how the power spectrum evolves with time on scales comparable to the free-streaming scale. Unlike other effects the suppression produced by neutrinos is redshift dependent, and by measuring at different redshifts it is possible to disentangle the effect of massive neutrinos from other effects coming from e.g.\ dark energy.

This type of measurement can be achieved with the LSST because the redshift of source galaxies can be determined reasonably well from photometry. It is therefore possible to do tomographical measurements of strucuture formation. It has been estimated that using data from Planck and LSST it will be possible to probe the sum of neutrino masses down to below 0.1 eV at 95\% C.L., i.e.\ it will be possible to tell the difference between the normal and the inverted hierarchy.
Table \ref{tab:tomo} shows the huge increase in sensitivity which can be achieved by doing weak lensing tomography. By binning the LSST data an improvement on $\sigma (\sum m_\nu)$ of almost a factor 4 can be achieved.
Other large scale structure surveys in the coming decade may reach a comparable sensitivity.

On the other hand, improving CMB measurements beyond Planck will not have a big impact on the determination of $m_\nu$ because in the context of $m_\nu$ CMB data are mainly used to break degeneracies with other parameters. For $N_\nu$ this is not the case because a change in $N_\nu$ has a very direct impact on the CMB spectrum. This is caused mainly by the change in the time of matter-radiation equality. From a detailed analysis Ref.~\cite{laurence} estimated the sensitivity of Planck to be $\sigma(N_\nu) \sim 0.3$ in a very general 11-parameter model, almost an order of magnitude better than WMAP.

\begin{table}
\caption{Estimated 1$\sigma$ sensitivity to $m_\nu$ from Planck and LSST with either no tomography or 5 tomographic bins \cite{tomography}.}
\label{tab:tomo}
\centering
\begin{tabular}{ll}\hline\hline
Data & $\sigma(\sum m_\nu)$ \\  \hline
Planck alone & 0.48 eV \\
Planck + LSST (no tomopraphy) & 0.15 eV \\
Planck + LSST (5 bins) & 0.043 eV \\ \hline \hline
\end{tabular}
\end{table}

\subsection{Non-linear structure formation and neutrinos}

Many of the upcoming precision experiments will probe structure formation in the quasi-linear regime $0.1 \, h \, {\rm Mpc}^{-1} < k < 1 \, h \, {\rm Mpc}^{-1}$, and to fully exploit their potential it is necessary to calculate observables
in this range to the required precision. In terms of the power spectrum this is $\sim 1-2$\% for $k > 1 \, h \, {\rm Mpc}^{-1}$.

\begin{figure}[h!]
\hspace{25mm}
\includegraphics[width=0.8\linewidth]{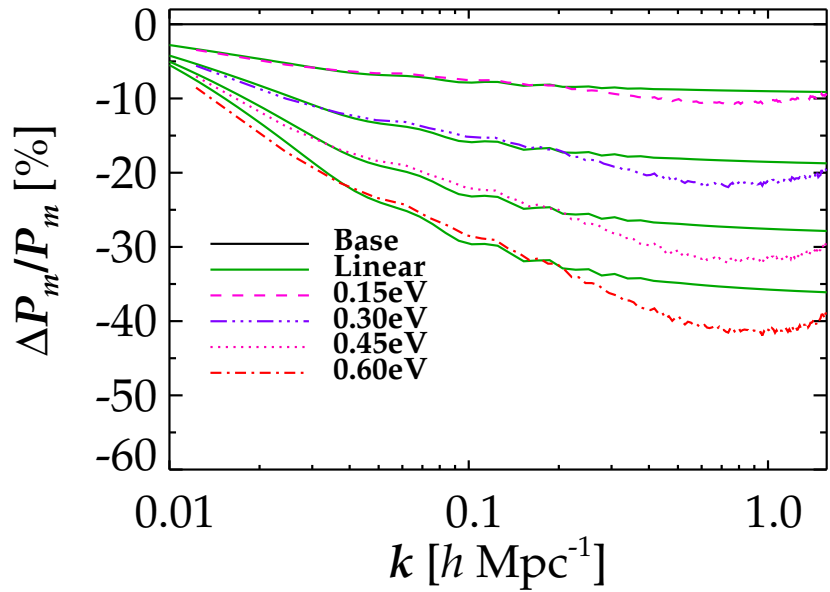}
\caption{The suppression of fluctuation power for different neutrino masses, compared with the linear theory result (taken from Ref.~\cite{brandbyge}).
\label{fig:damping}}
\end{figure}

N-body simulations of non-linear structure formation with neutrinos included have shown that neutrinos provide an additional, non-linear suppression of power beyond the usual linear theory $\Delta P/P \sim 8 \Omega_\nu/\Omega_m$.
Fig.~\ref{fig:damping} shows the power spectrum suppression for various values of the neutrino mass.
Very interestingly, the damping relative to a pure $\Lambda$CDM model is larger than in linear theory, an effect caused by the interplay between the neutrino thermal velocity and the gravitational virial velocity \cite{brandbyge}. The maximum relative suppression (which does not occur at asymptotically high $k$) is given approximately by
\begin{equation}
\frac{\Delta P}{P} \sim -9.6 f_\nu.
\end{equation}
This additional suppression, which is unique to neutrinos, might provide a smoking gun signature for the presence of massive neutrinos. It will be almost impossible to mimic with other physics because it requires the presence of a second dark matter component with very specific thermal properties.

\section{Measuring the relic neutrino background}

Standard model
physics likewise predicts the presence of a Cosmic Neutrino Background (\cnub) with a well defined
temperature of $T_\nu \sim (4/11)^{1/3} T_\gamma$. While it remains undetected in direct experiments, the
presence of the {\cnub} is strongly hinted at in CMB data. The homogeneous C$\nu$B component has been
detected at the 4-5$\sigma$ level in the WMAP data (see e.g.\ \cite{Komatsu:2008hk,Hamann:2007pi,deBernardis:2007bu,Ichikawa:2008pz,Hamann:2008we,Popa:2008tb}). Furthermore, this component is known to be
free-streaming, i.e. to have an anisotropic stress component consistent with what is expected from standard
model neutrinos (see
\cite{Bashinsky:2003tk,Trotta:2004ty,Bell:2005dr,DeBernardis:2008ys,Basboll:2008fx,Hannestad:2004qu,Friedland:2007vv}).
Finally the standard model neutrino decoupling history is also confirmed by Big Bang Nucleosynthesis (BBN),
the outcome of which depends on both the energy density and flavour composition of the {\cnub}.

While this indirect evidence for the presence of a {\cnub} is important, a direct detection remains an
intriguing, but almost impossible goal. The most credible proposed method is to look for a peak in beta
decay spectra related to neutrino absorption from the {\cnub} \cite{Weinberg:1962zz,Cocco:2007za,Blennow:2008fh}, although
many other possibilities have been discussed
\cite{Weiler:1982qy,Stodolsky:1974aq,Gelmini:2004hg,Ringwald:2004np,Fodor:2002hy,Duda:2001hd,Langacker:1982ih,Cabibbo:1982bb}.
The neutrino absorption method was first investigated by Weinberg \cite{Weinberg:1962zz}, based on the
possibility that the primordial neutrino density could be orders of magnitude higher than normally assumed
due to the presence of a large chemical potential. Although a large chemical potential has been ruled out
because it is in conflict with BBN and CMB
\cite{Pastor:2001iu,Pastor:2008ti,Simha:2008mt,Wong:2002fa,Abazajian:2002qx}, the method may still work and
recently there has been renewed interest in detecting the {\cnub} using beta unstable nuclei.

Although the direct detection of the {\cnub} is already very challenging, one might speculate on the
possibility that in the more distant future anisotropies in the {\cnub} will be detectable.

\subsection{Anisotropy of the background}

\begin{figure}
   \noindent
      \begin{center}
      \hspace*{-0.1cm}\includegraphics[width=0.8\linewidth]{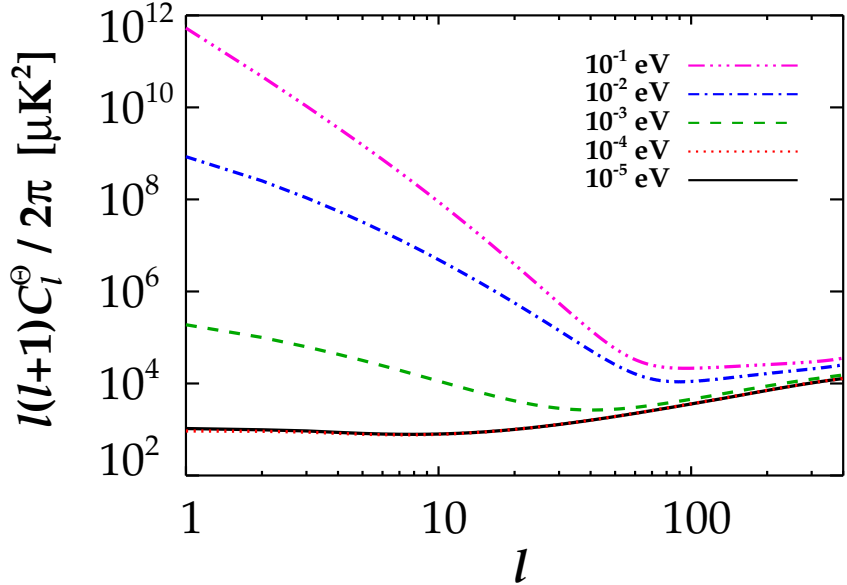}
      \end{center}
   \caption{Primary {\cnub} spectrum for different neutrino masses \cite{sth09}.}
   \label{fig:cnub}
\end{figure}

In Ref.~\cite{sth09} the anisotropy of the {\cnub} was calculated, and the result is shown in Fig.~\ref{fig:cnub}. For massive neutrinos the anisotropy increases dramatically on large scales and reaches $\cal{O}$(1) for a mass of order 0.05-0.1 eV. Assuming that the {\cnub} anisotropy is to be detected using a beta decay experiment the actual measured anisotropy will be a linear combination of the mass eigenstates, weighed with their mixing with $\nu_e$. Since the anisotropy grows very strongly with increasing mass, the main signal will almost solely come from the most massive eigenstate, even if it has very small mixing.

\section{Conclusions}

Cosmology remains one of the main tools for the study of neutrino physics. Currently cosmology provides the most stringent upper bound on the neutrino mass, and even though the exact number is model dependent a very conservative upper bound on the sum of neutrino masses can be put at $\sum m_\nu \lwig 0.6-0.7$ eV at 95\% C.L. More aggressive use of data leads to more stringent bounds but in that case it is also necessary to rely on less well controlled effects from non-linear structure formation.
The other important parameter which can be probed using structure formation data is the energy density in neutrinos at early times, quantified by the parameter $N_\nu$, the effective number of neutrino species.
The current bound from the same set of data is $3.03 < N_\nu < 7.59$ at 95\% C.L.
Intriguingly, the preferred value of $N_\nu$ is consistently higher than the standard model value of 3.04, but not at more than approximately 2$\sigma$.

In the future a range of different experiments will improve the sensitivity to neutrino parameters. Most important for a precision determination of $\sum m_\nu$ will be measurements of the matter power spectrum using larger volumes and going to higher redshifts than current surveys. During the next decade the most precise data will probably come from weak lensing survey of the LSST telescope. Together with measurements of the CMB anisotropy by the Planck satellite it has been estimated that a 1$\sigma$ uncertainty on $\sum m_\nu$ of $\sim 0.04$ eV can be achieved.
In the more distant future it may be possible to decrease this error bar significantly by large scale measurements of 21-cm fluctuations at very high redshift, using for example the proposed FFTT project \cite{fftt}. This could in principle increase the sensitivity by another factor of a few, making a precise neutrino mass determination possible even for the normal hierarchy. However, it should be stressed that there are many currently unadressed systematics involves in this, and it remains unclear if 21-cm surveys can ever reach this sensitivity.

In conclusion, cosmology is an important and complementary laboratory for probing neutrino physics. Some neutrino parameters, like the neutrino mass, are in principle much easier to measure using precision cosmological data than in direct laboratory experiments. Furthermore, since cosmology is measuring a different effective mass quantity than beta or double beta decay experiments it remains an intriguing possibility that they will yield different and seemingly incompatible results. For example it may be the case that cosmology provides a stringent upper limit while for example a beta decay experiment shows positive evidence for a non-zero $m_\nu$. Such a possibility could point to non-standard physics such as right handed currents masking as a neutrino mass in the beta decay experiment.

\end{document}